\documentclass[a4paper,12pt]{article}

\newcommand{\be}{\begin{equation}}
\newcommand{\ee}{\end{equation}}
\newcommand{\bea}{\begin{eqnarray}}
\newcommand{\nn}{\nonumber}
\newcommand{\eea}{\end{eqnarray}}

\begin{document}

\begin{titlepage}
\begin{flushright}
UB-ECM-PF-03/04
\end{flushright}
\begin{centering}
\vspace{.3in}
{\Large{\bf Generalization of the KKW Analysis\\ for\\ Black Hole
Radiation}}
\\

\vspace{.5in} {\bf  Elias C.
Vagenas\footnote{evagenas@ecm.ub.es} }\\

\vspace{0.3in}

Departament d'Estructura i Constituents de la Mat\`{e}ria\\
and\\ CER for Astrophysics, Particle Physics and Cosmology\\
Universitat de Barcelona\\
Av. Diagonal 647\\ E-08028 Barcelona\\
Spain\\
\end{centering}

\vspace{0.6in}
\begin{abstract}
\noindent An extension of the Keski-Vakuri, Kraus and Wilczek
(KKW) analysis to black hole spacetimes which are not
Schwarzschild-type is presented. Preserving the regularity at the
horizon and stationarity of the metric in order to deal with the
across-horizon physics, a more general coordinate transformation
is introduced. In this analysis the Hawking radiation is viewed as
a tunnelling process which emanates from the
non-Schwarzschild-type black hole. Expressions for the temperature
and entropy of these non-Schwarzschild-type black holes are
extracted. As a paradigm, in the context of this generalization,
we consider the Garfinkle-Horowitz-Strominger (GHS) black hole as
a dynamical background and  we derive the modified temperature and
entropy of GHS black hole. Deviations are eliminated and
corresponding standard results are recovered to the lowest order
in the emitted shell of energy. The extremal GHS black hole is
found to be non-``frozen'' since it is characterized by a constant
non-zero temperature. Furthermore, the modified extremality
condition forbids naked singularities to form from the collapse of
the GHS black hole.
\end{abstract}
\end{titlepage}

\newpage

\baselineskip=18pt
\section*{Introduction}
\hspace{0.8cm} The idea of Keski-Vakkuri, Kraus and Wilczek
\cite{kraus1,kraus2,keski} (KKW) has been applied till now to
Schwarzschild-type black hole geometries. In this semiclassical
analysis the energy conservation plays a dominant role since it
leads us to a dynamical Schwarzschild-type black hole background
which in turn leads to a more realistic description of the black
hole radiance \cite{steve,hawking}. The total
Arnowitt-Deser-Misner mass \cite{adm} is fixed while the mass of
the Schwarzschild-type black hole decreases due to the emitted
radiation. A coordinate transformation is implemented so that the
line element be non-singular at the horizon. This permits the
study of across-horizon physics such as the black hole radiation.
Using this methodology, it is has been made possible to exactly
evaluate the non-thermal spectrum of the black hole radiation (the
non-thermal character of the black hole radiation was already
known since the seminal work of Hawking in mid-seventies). A
direct byproduct of this analysis is that the black hole
temperature is not only a function of the characteristics of the
black hole but also of the energy of the emitted shell of energy.
Additionally the black hole entropy is not that given by the area
formula of Bekenstein and Hawking for the corresponding
Schwarzschild-type black hole.
\par
In the seminal works of  Keski-Vakkuri, Kraus and Wilczek
\cite{kraus1,kraus2,keski}, although the starting point was
spherically symmetric geometries, the analysis was restricted to
Schwarzschild-type black holes, i.e. no general expressions for
temperature and entropy were extracted for the above-mentioned
geometries. Here we introduce a more general coordinate
transformation in order to apply the KKW analysis to
non-Schwarzschild-type black hole spacetimes. This general
transformation preserves two conditions: (a) the regularity at the
horizon which ensures that we are able to study across-horizon
physics, (b) the stationarity of the non-static metric which
implies that the time direction is a Killing vector, which are
crucial in order to generalize the KKW analysis. The methodology
followed is an appropriate modification of that adapted in
 the Schwarzschild-type black holes.
The effect of this generalization in our calculation leads to
exact expressions for the temperature and the entropy of  the
non-Schwarzschild-type black holes which are not anymore the
Hawking temperature and the Bekenstein-Hawking entropy (given by
the area formula), respectively.
\par
The outline of this paper is as follows. Section 1 is devoted to
the presentation of KKW analysis in Schwarzschild-type black hole
geometries. In Section 2  we extend the analysis of Section 1 to
the case of non-Schwarzschild-type black hole geometries. In the
framework of this semiclassical analysis, we derive exact
expressions for the temperature and entropy of these black hole
spacetimes. In Section 3 we implement the above-mentioned
expressions for the case of the GHS black hole. We derive the
corresponding modified temperature and entropy of the GHS black
hole and in the lowest order of the emitted shell of energy
standard results, i.e. Hawking temperature and Bekenstein-Hawking
entropy are reproduced, respectively, as a verification of the
validity of our results. We consider the extremality condition
which now will be shifted since the charge $Q$ of the
four-dimensional GHS black hole will be reached by the mass $M$
earlier. The temperature of the extremal GHS black hole is shown
to be non-zero and singularities are found to be always hidden
behind the event horizon of the GHS black hole. Finally, in
Section 4 we end up with a short summary and concluding remarks.
\section{KKW Analysis}
The idea of Keski-Vakkuri, Kraus and Wilczek (KKW) was firstly
utilized for the case of the four-dimensional Schwarzschild black
hole by Parikh and Wilczek \cite{parikh}. From that time till now
the KKW methodology was also applied to several black hole
solutions such as (d+1)-dimensional Anti-de-Sitter black hole
\cite{hemming}, AdS(2) black hole \cite{kwon}, two-dimensional
charged black hole solution derived from the effective string
theory at the low-energy limit \cite{elias1}, two-dimensional
charged (and uncharged) dilatonic black holes \cite{elias2}
(dimensionally reduced from (2+1) spinning (and spinless) BTZ
black holes), (2+1)-dimensional charged BTZ black hole
\cite{elias3,medved1} and lately to a Schwarzschild-de Sitter
spacetime \cite{medved2}.
\par\noindent
All these black hole backgrounds belong to the same family of
geometries since their line elements were Schwarzschild-like, i.e.
they were of the type \be ds^2 = -A(r)dt^2 + A^{-1}(r)dr^2 +
r^{2}d\Omega \label{linelement} \ee where the metric function
$A(r)$ had at least one (outer) event  horizon ($r_{+}$), i.e. \be
A(r_{+})=0\hspace{1ex}.\ee
\par\noindent
Since the frame of the KKW methodology is the Hawking phenomenon which main
contribution comes from the event horizon, the line element should be regular at the
event horizon. Therefore a choice of suitable coordinates was enforced and the
ansantz is to pick up the $a\hspace{1ex}l\acute{a}$ Painlev$\acute{e}$  \cite{painleve} coordinate
transformation \be \sqrt{A(r)}\;dt = \sqrt{A(r)}\;d\tau -\sqrt{\frac{1-A(r)}{A(r)}}\;dr
\label{painleve} \ee where $\tau$ is the new time coordinate (Painlev$\acute{e}$
coordinate).
\par\noindent After squaring expression (\ref{painleve}) and substituting
into equation (\ref{linelement}), the line element becomes \be
ds^2 = -A(r)d\tau^2 + 2\sqrt{1-A(r)}d\tau dr + dr^2 +
r^{2}d\Omega\hspace{1ex}. \ee It is obvious that there is no
singularity at the event horizon $r_H$ and these coordinates are
stationary, but not static.
\par\noindent
The radial null geodesics are given by \be
\dot{r}\equiv\frac{dr}{d\tau}=\pm 1-\sqrt{1-A(r)}\label{geod1}\ee
where the upper (lower) sign in the above equation corresponds
(under the assumption that $\tau$ increases towards future) to the
outgoing (ingoing) geodesics.\par\noindent At this point we will
take into consideration the self-gravitation effect by fixing the
total Arnowitt-Deser-Misner mass ($M_{ADM}$) of the black hole and
letting the mass $M$ of the black hole to vary. A shell of energy
$\omega$ is now radiated by the black hole. It travels on the
outgoing geodesics which now are  due to the fluctuation of the
mass $M$ of the black hole, derived by the modified line element
\be ds^2 = -A(r, M-\omega)d\tau^2 + 2\sqrt{1-A(r, M-\omega)}d\tau
dr + dr^2 + r^{2}d\Omega\hspace{1ex}. \ee The outgoing radial null
geodesics followed by the shell of energy will also be modified
as\be\dot{r}=1-\sqrt{1-A(r,
M-\omega)}\label{geod2}\hspace{1ex}.\ee At this point let us
remind ourselves of the following statements
\begin{enumerate} \item It is known that the emission rate $\Gamma$ for a radiating
source \cite{hawking} is given as \be \Gamma\approx e^{-\beta
\omega}=e^{+\Delta S_{bh}}\ee where $\beta$ is the inverse
temperature ($T_{bh}$) of the black hole and $\Delta S_{bh}$ is
the change in the entropy of the black hole before and after the
emission of the shell of energy $\omega$ (outgoing massless
particle) \be\Delta S_{bh}=S_{bh}(M -
\omega)-S_{bh}(M)\hspace{1ex}.\ee \item A canonical Hamiltonian
treatment gives a simple result for the total action of a system
\cite{kraus1} \be \mathcal{I}=\int
d\tau\left[p_{\tau}+\frac{dr}{d\tau}p_{r}\right]\label{hamilton}\ee
where $\tau$ and $r$ are the Painlev$\acute{e}$ coordinate while
$p_{\tau}$ and $p_{r}$ are the corresponding conjugate momenta.
\item A semiclassical (WKB) approximation gives the following
expression for the emission rate \cite{keski} \be\Gamma\approx
e^{-2Im\mathcal{I}}\ee where only the second term in equation
(\ref{hamilton}) contributes to the imaginary part of the action.
\end{enumerate}
We will consider here only the s-wave emission of massless particles.
Therefore, using the above mentioned statements to the KKW methodology the imaginary
part of the action can be obtained \bea
Im\mathcal{I}&=&Im\int\frac{dr}{d\tau}\,p_r d\tau\\
&=&Im\int_{r_{+}(M)}^{r_{+}(M-\omega)}p_r dr\label{imaginary}\eea
where $r_{+}$ is the outer event horizon of the black hole as
mentioned before. It is useful to apply the Hamilton's equation
\bea\dot{r}&=&\frac{dH}{dp_{r}}\\&=&\frac{d(M-\omega)}{dp_{r}}\eea
and thus \be dp_{r}=\frac{d(M-\omega)}{\dot{r}}\label{moment} \ee
 Equation (\ref{imaginary}) can be  written as
 \be
 Im\mathcal{I}=Im\int_{r_{+}(M)}^{r_{+}(M-\omega)}\int^{p_r}_{0}dp_{r}'
  dr\label{imaginary1}\ee and substituting equation (\ref{moment}) we get \be
  Im\mathcal{I}=Im\int_{r_{+}(M)}^{r_{+}(M-\omega)}\int^{+\omega}_{0}
  \frac{d(M-\omega')}{\dot{r}}dr\label{imaginary2}\hspace{1ex}.\ee Thus if we use the
  modified outgoing geodesics (\ref{geod2}) the imaginary part of the action
  will be written as
  \be Im\mathcal{I} =Im\int_{r_{+}(M-\omega)}^{r_{+}(M)}\int^{+\omega}_{0}
  \frac{d\omega'}{1-\sqrt{1-A(r,
M-\omega')}}dr\label{imaginary3}\hspace{1ex}.\ee It is easily seen, using the
first statement, that the temperature and the entropy of the black hole is not
the Hawking temperature ($T_H$) and the  entropy given by the area formula of Bekenstein and
Hawking ($S_{BH}$), respectively. Both of them are  modified due to the specific modelling
of the self-gravitation effect.
Thus, the modified temperature and entropy of the black hole is given, respectively, by
\be
T_{bh}=\frac{\omega}{2}\left\{Im\int_{r_{+}(M-\omega)}^{r_{+}(M)}\int^{+\omega}_{0}
  \frac{d\omega'}{1-\sqrt{1-A(r,
M-\omega')}}dr\right\}^{-1}
\label{mtemp}
\ee
\be
S_{bh}(M-\omega)=S_{bh}(M)
-2Im\int_{r_{+}(M-\omega)}^{r_{+}(M)}\int^{+\omega}_{0}
  \frac{d\omega'}{1-\sqrt{1-A(r,
M-\omega')}}dr
\ee
where the entropy of the black hole with mass $M$ must be
equal to that given by the area formula of Bekenstein and
Hawking ($S_{BH}$). Therefore, the expression for
the modified entropy will be
\be
S_{bh}(M-\omega)=S_{BH}
-2Im\int_{r_{+}(M-\omega)}^{r_{+}(M)}\int^{+\omega}_{0}
  \frac{d\omega'}{1-\sqrt{1-A(r,
M-\omega')}}dr
\label{mentrop}\hspace{1ex}.
\ee
\section{Generalizing the KKW Analysis}
The KKW methodology was,  till now, restricted to
Schwarzschild-type black hole backgrounds (see equation
(\ref{linelement})). In this section we extend the KKW analysis to
more general black hole backgrounds of the form \be ds^2 =
-A(r)dt^2 + B^{-1}(r)dr^2 + r^{2}d\Omega \label{glinelement} \ee
where $A(r)$ and $B(r)$ are functions satisfying the equation \be
A(r)\cdot B^{-1}(r)\not= 1\hspace{1ex}. \ee The metric function
$B(r)$ had at least one (outer) event  horizon ($r_{+}$), i.e. \be
B(r_{+})=0\ee since a event horizon appears at a spacetime point
where $g^{rr}=0$ and for the line element in (\ref{glinelement})
\be g^{rr}=B(r)\hspace{1ex}.\ee In order for the total
Arnowitt-Deser-Misner mass ($M_{ADM}$) to be well-defined we
restrict the class of metrics to those which are asymptotically
flat, i.e. \bea
A(r)\rightarrow 1 & \mbox{as} &  r\rightarrow +\infty \nn\\
B(r) \rightarrow 1& \mbox{as} & r\rightarrow
+\infty\nn\hspace{1ex}. \eea
\par\noindent
It is obvious that since we would like to deal with the radiation
phenomenon of a black hole we need to keep the regularity at the
event horizon and the stationarity of the metric which in turn
implies that the time direction is a Killing vector
\cite{per,maulik}. Therefore, we introduce the following
$a\hspace{1ex}l\acute{a}$ Painlev$\acute{e}$  more general,
compared to that applied before for the Schwarzschild-type black
holes, coordinate transformation \be \sqrt{A(r)}\;dt =
\sqrt{A(r)}\;d\tau -\sqrt{B^{-1}(r)-1}\;dr \label{gpainleve} \ee
where $\tau$ is the new time coordinate (Painlev$\acute{e}$
coordinate). Substituting expression (\ref{gpainleve}) in equation
(\ref{glinelement}) the line element becomes \be ds^2 =
-A(r)d\tau^2 + 2\sqrt{\frac{A(r)}{B(r)}\bigg(1-B(r)\bigg)}\,d\tau
dr + dr^2 + r^{2}d\Omega\hspace{1ex}. \ee
\par\noindent
The radial null geodesics are now given by \be
\dot{r}=\sqrt{\frac{A(r)}{B(r)}}\left[ \pm
1-\sqrt{1-B(r)}\;\right] \label{ggeod1} \ee where the upper
(lower) sign in the above equation corresponds, as before, to the
outgoing (ingoing) geodesics under the assumption that $\tau$
increases towards future.
\par\noindent
At this point we fix the total Arnowitt-Deser-Misner mass
($M_{ADM}$) of the black hole since we want to include the effect
of self-gravitation. On the contrary, we let the mass $M$ of the
black hole to fluctuate. A shell of energy $\omega$ which
constitutes of massless particles considering only the s-wave part
of emission, is now radiated by the black hole. The massless
particles travel on the outgoing geodesics which now are  due to
the varying mass $M$ of the black hole, derived by the modified
line element \be ds^2 = -A(r, M-\omega)d\tau^2 + 2\sqrt{\frac{A(r,
M-\omega)}{B(r, M-\omega)}\bigg(1-B(r, M-\omega)\bigg)}d\tau dr +
dr^2 + r^{2}d\Omega\hspace{1ex}.\ee The outgoing radial null
geodesics followed by the massless particles, i.e. the shell of
energy, will also be modified as follows \be
\dot{r}=\sqrt{\frac{A(r, M-\omega)}{B(r, M-\omega)}}\left[ \pm
1-\sqrt{1-B(r, M-\omega)}\;\right] \label{ggeod2}\hspace{1ex}. \ee
We adopt the previously mentioned three statements. We follow the
same steps as before in order to write down an expression for the
imaginary part of the action. Using the modified outgoing
geodesics (\ref{ggeod2}), the imaginary part of the action will
now be given by \be Im\mathcal{I}
=Im{\int_{r_{+}(M-\omega)}^{r_{+}(M)}}\int^{+\omega}_{0}
  \frac{d\omega'}{\sqrt{\frac{\widetilde{A}}{\widetilde{B}}}\left[1-\sqrt{1-\widetilde{B}}\;\right]}dr
\label{gimaginary3}
\ee
where $\widetilde{A}$ and $\widetilde{B}$ are defined as follows
\bea
\widetilde{A}&=&A(r, M-\omega')\\
\widetilde{B}&=&B(r, M-\omega')\hspace{1ex}.
\eea
\par\noindent
Finally considering a more general black hole geometry than that of the Schwarzschild-type,
the expression for the modified temperature is given as
\be
T_{bh}=\frac{\omega}{2}\left\{Im\int_{r_{+}(M-\omega)}^{r_{+}(M)}\int^{+\omega}_{0}
  \frac{d\omega'}{\sqrt{\frac{\widetilde{A}}{\widetilde{B}}}\left[1-\sqrt{1-\widetilde{B}}\;\right]}dr
\right\}^{-1}
\label{gmtemp}
\ee
and the corresponding expression for the modified entropy is given as
\be
S_{bh}(M-\omega)=S_{BH}
-2Im\int_{r_{+}(M-\omega)}^{r_{+}(M)}\int^{+\omega}_{0}
  \frac{d\omega'}{\sqrt{\frac{\widetilde{A}}{\widetilde{B}}}\left[1-\sqrt{1-\widetilde{B}}\;\right]}dr
\label{gmentrop}\hspace{1ex}. \ee It is obvious that in case where
the metric functions $A$ and  $B$ satisfy the condition \be A(r)
\cdot B^{-1}(r) =1\, , \ee
 equations (\ref{gmtemp}) and (\ref{gmentrop}) coincide with the respective expressions
(\ref{mtemp}) and (\ref{mentrop}) for the Schwarzschild-type black hole.
\section{GHS Black Hole}
The starting point will be the four-dimensional low-energy action
obtained from string theory which is written in terms of the
string metric as \cite{ghs} \be S=\int d^{4}x
\sqrt{-g}\;e^{-2\phi}\left[-R-4\left(\nabla
\phi\right)^{2}+F^2\right] \ee and the charged black hole metric
is \be
ds^{2}_{\tiny\mbox{string}}=-\frac{\left(1-\frac{2Me^{\phi_0}}{r}\right)}
{\left(1-\frac{Q^{2}e^{3\phi_0}}{M r }\right)}dt
^{2}+\frac{dr^{2}}
{\left(1-\frac{2Me^{\phi_0}}{r}\right)\left(1-\frac{Q^{2}e^{3\phi_0}}
{Mr}\right)}+ r^{2}d\Omega\hspace{1ex}.\label{ghsmetric}\ee This
metric describes a black hole with an event horizon at \be r_{+}=2
M e^{\phi_0} \ee when $Q^{2}<2e^{-2\phi_{0}}M^{2}$. The Hawking
temperature of the GHS black hole solution (\ref{ghsmetric})
easily evaluated by the use of the periodicity of the Euclidean
section, is given by \be T_{H}=\frac{1}{8\pi M
e^{\phi_{0}}}\label{ghstemp}\hspace{1ex}.\ee It is obvious that
the Hawking temperature of the GHS black hole is independent of
the charge $Q$, for $Q<\sqrt{2}e^{-\phi_{0}}M$. \par\noindent At
extremality, i.e. when $Q^{2}=2e^{-\phi_{0}}M^{2}$, the  GHS black
hole solution (\ref{ghsmetric}) becomes \be
ds^{2}_{\tiny\mbox{string}}=- dt^{2}+
\left(1-\frac{2Me^{\phi_0}}{r}\right)^{-2}dr^{2}+
r^{2}d\Omega\hspace{1ex}.\label{extghsmetric}\ee and the
corresponding Hawking temperature of the extremal GHS black hole
(\ref{extghsmetric}) is \be
T_{H}^{\mbox{\tiny{ext}}}=0\label{extghstemp}\ee since the
Euclidean section is smooth but without
identifications.\par\noindent In order to implement the
methodology introduced in the previous section we firstly identify
the metric functions $A(r)$ and $B(r)$ for the case of GHS black
hole by comparing equation (\ref{glinelement}) with
(\ref{ghsmetric}) and we get \bea A(r)&=&
\frac{\left(1-\displaystyle{\frac{2Me^{\phi_0}}{r}}\right)}
{\left(1-\displaystyle{\frac{Q^{2}e^{3\phi_0}}{Mr}}\right)}\label{ghs1}\\B(r)&=&\left(1-\frac{2Me^{\phi_0}}
{r}\right)\left(1-\frac{Q^{2}e^{3\phi_0}}
{Mr}\right)\label{ghs2}\hspace{1ex}.\eea Avoiding to make the
complicated computation of the integral in expression
(\ref{gimaginary3}) for the specific black hole background
(\ref{ghsmetric}) we make the following approximation \be
\sqrt{\frac{\widetilde{A}}{\widetilde{B}}}\left(1-\sqrt{1-\widetilde{B}}\;\right)
\approx\frac{1}{2}\sqrt{\widetilde{A}\widetilde{B}}\label{approx}\hspace{1ex}.\ee
Substituting expression (\ref{approx}) in equation
(\ref{gimaginary3}) and using the expressions of $A(r)$ and $B(r)$
for the GHS black hole, i.e. equations (\ref{ghs1}) and
(\ref{ghs2}), respectively, we get \bea Im\mathcal{I}&\approx& 2\,
Im{\int_{r_{+}(M-\omega)}^{r_{+}(M)}}\int^{+\omega}_{0}
\frac{d\omega'dr}{\sqrt{\widetilde{A}\widetilde{B}}}\nn\\
&=& 2\,Im\int_{r_{+}(M-\omega')}^{r_{+}(M)}
\int^{+\omega}_{0}\frac{d\omega'
dr}{\left(1-\displaystyle{\frac{2(M-\omega)e^{\phi_0}}{r}}\right)}\hspace{1ex}.\eea
We firstly perform the $\omega$-integration which involves a
contour integration into the lower half of $\omega'$ plane and we
finally get \be
Im\mathcal{I}=\frac{\pi}{2}e^{-\phi_{0}}\left[r^{2}_{+}(M)-
r^{2}_{+}(M-\omega)\right]\label{approximag}\hspace{1ex}. \ee Thus
the modified temperature (\ref{gmtemp}) for the case of the GHS
black hole (\ref{ghsmetric}) is given as \be T_{bh}(M, \phi_{0},
\omega)= \frac{\omega}{4\pi M^{2}e^{\phi_{0}}}\left
[1-\left(1-\frac{\omega}{M}\right)^{2}\right]^{-1}\label{mghstemp}\ee
and the corresponding expression for the modified entropy
(\ref{gmentrop}) is given as \be S_{bh}(M, \phi_{0},
\omega)=S_{BH}-4\pi
M^{2}e^{\phi_{0}}\left[1-\left(1-\frac{\omega}{M}\right)^{2}\right]\label{mghsentropy}\hspace{1ex}.\ee
We see that there are deviations from the standard results derived
for a fixed background. The temperature of the GHS black hole is
not the Hawking temperature (\ref{ghstemp}) and its entropy is not
given by  the Bekenstein-Hawking area formula \cite{cad} \bea
S_{BH}&=&\frac{1}{4}\mathcal{A}_{H}\nn\\
&=&\pi r_{+}^{2}=4\pi
M^{2}e^{2\phi_{0}}\label{bhentropy}\hspace{1ex}.\eea A welcomed
but not unexpected result is that the modified temperature
(\ref{mghstemp}) evaluated to first order in $\omega$  yields the
Hawking temperature of the GHS black hole (\ref{ghstemp}).
Additionally the modified entropy of the GHS black hole
(\ref{mghsentropy}) to zeroth order in $\omega$ yields the
corresponding Bekenstein-Hawking entropy (\ref{bhentropy}).
\par\noindent In the framework of our analysis, the extremal GHS black hole will be created when
\be Q^{2}=2e^{-\phi_{0}}(M-\omega)^{2}\hspace{1ex}.\ee It is
obvious that the extremality condition ($r_{+}=r_{-}$) is modified
and the temperature of the extremal GHS black hole is no longer
zero but it is given as \be T^{\mbox{\tiny{ext}}}_{bh}(M, Q,
\phi_{0} )=\frac{1}{4\pi M e^{-\phi_{0}}
\left(1-\displaystyle{\frac{Q}{\sqrt{2}M}e^{-\phi_{0}}}\right)}\hspace{1ex}.\ee
As a byproduct of this modification to the extremality condition,
since the emitted shell of energy $\omega$ has to be always
positive \be
\omega=M-\frac{Q}{\sqrt{2}}e^{\phi_{0}}>0\hspace{1ex},\ee it is
implied that \be Q < \sqrt{2} M e^{-\phi_{0}}\hspace{1ex}. \ee
Thus the extremality condition indicates that a naked singularity
will never form from the collapse of the GHS black hole.
\section{Conclusions}
We have introduced a new, more general, coordinate transformation
in order to generalize the KKW analysis to non-Schwarzschild-type
black holes. Exact expressions for the temperature and the entropy
of these black holes have been derived. Due to the specific
modelling of the self-gravitation effect - described by the KKW
analysis - the black hole temperature is not the Hawking
temperature but depends explicitly on the emitted masssless (since
we have restricted our analysis to the s-wave emission) particle's
energy. The black hole entropy is also different from the
corresponding entropy given by the area formula of Bekenstein and
Hawking. It is easily seen that the modified entropy is less than
the Bekenstein-Hawking entropy. In the context of our generalized
KKW analysis, we have implemented the aforesaid expressions for
the case of the static four-dimensional charged black holes in
string theory which are the well-known
Garfinkle-Horowitz-Strominger (GHS) black holes. The temperature
of the GHS black hole is no more the corresponding Hawking
temperature and the entropy of the GHS black hole is no longer the
corresponding Bekenstein-Hawking entropy. The ``greybody factors''
showing up declare explicitly the dependence on the emitted
particle's energy. As a verification of the validity of our
generalized KKW analysis introduced here, the modified temperature
and entropy of GHS black hole in the lowest order of the emitted
shell of energy reproduce the standard results. Finally, we have
shown that the extremal GHS black hole is no more ``frozen'' but
it is characterized by  a nonzero background temperature since the
extremality condition is modified (due to the specific modelling
of the backreaction effect). It should also be noted that because
of the modified extremality condition naked singularities are
forbidden in a natural way.
\section*{Acknowledgements}
The author would like to thank the anonymous referee of Physics
Letters B for his useful comments and Ass. Professor T.
Christodoulakis for reading the manuscript.  This work has been
supported by the European Research and Training Network
``EUROGRID-Discrete Random Geometries: from Solid State Physics to
Quantum Gravity" (HPRN-CT-1999-00161).

\end{document}